# Quantum electrical transport properties of topological insulator $Bi_2Te_3$ nanowires

Hong-Seok Kim,[1] Ho Sun Shin,[2] Joon Sung Lee,[3]* Chi Won Ahn,[4] Jae Yong Song,[2] Yong-Joo Doh,[1]*

[1]*Department of Applied Physics, Korea University Sejong Campus, Sejong 339-700, Republic of Korea*

[2]*Center for Nanomaterials Characterization, Korea Research Institute of Standards and Science, Daejeon 305-340, Republic of Korea*

[3]*Department of Display and Semiconductor Physics, Korea University Sejong Campus, Sejong 339-700, Republic of Korea*

[4] *Nano-Materials Laboratory, National Nanofab Center, Daejeon 305-806, Republic of Korea*

*e-mail: simmian@korea.ac.kr, yjdoh@korea.ac.kr

**Abstract**

We investigate the quantum transport properties of surface electrons on a topological insulator $Bi_2Te_3$ nanowire in a magnetotransport study. Although the nanowires are synthesized by using a relatively coarse method of electrochemical deposition, clear Aharonov–Bohm oscillations of phases 0 and π are observed, owing to the highly coherent surface electron channel. The oscillation amplitude exhibits exponential temperature dependence, suggesting that the phase coherence length $L_\varphi$ is inversely proportional to the temperature, as in quasi-ballistic systems. In addition, a weak antilocalization analysis on the surface channel by using a one-dimensional localization theory, enabled by successful extraction of the surface contribution from the magnetoconductance data, is provided in support of the temperature dependence of $L_\varphi$.

## 1. Introduction

Topological insulators (TIs) exhibit a novel gapless spin-resolved edge (or surface) electronic state[1-2] that is located within the bulk insulating energy gap, with spin−momentum locking due to a strong spin−orbit interaction. The TI system is highly promising not only for fundamental studies on topological nature in condensed matter physics, but also for potential applications in future information technologies such as topological quantum computing and spintronics.[1-4] A two-dimensional (2D) topological insulating state was theoretically predicted for HgTe quantum well structures,[3] and experimentally verified by using a transport study.[5] Later, $Bi_{1-x}Sb_x$ alloys were determined to be three-dimensional (3D) TIs[4] with a topological surface state (TSS) by an angle-resolved photoemission spectroscopy (ARPES) study.[6] Recently, additional materials, such as $Bi_2Se_3$, $Bi_2Te_3$, and $Sb_2Te_3$, have also been identified as 3D TIs, of which the surfaces have single Dirac cone band structures protected by time-reversal symmetry.[7-9] In addition, the existence of a spin-helical state, in which the electron spin is aligned parallel to the surface and normal to the momentum, has been verified from spin-resolved ARPES studies.[8, 10]

TIs may sustain highly coherent charge- and spin-transport by virtue of the spin−momentum helical locking and the protection from backscattering by time reversal symmetry in the TSS; hence, they can constitute an attractive platform for spintronic device applications. Further progress in developing TI-based applications needs to be preceded by experimental studies on their electrical transport characteristics for determining the surface state properties. However, the contribution from the residual bulk carriers due to material imperfection makes it difficult to isolate the genuine transport properties of the TSS. This issue has been tackled with methods such as tuning the Fermi level to the bulk band gap by chemical doping[11-13] or external gate control,[14-15] and by using nanostructures of TIs, which cause the surface state contribution to dominate the charge transport, owing to the large surface-to-volume ratio.[16-18]

Conduction electrons in the TSS of TIs exhibit strong quantum interference effects. Weak antilocalization (WAL)[19] occurs in TIs as a result of a strong spin-orbit interaction, which is known to disturb the weak localization effect caused by constructive interference between time-reversed electron paths in diffusive media. Analyses of magnetoconductance (MC) data for WAL have been performed to investigate the phase coherence and spin−orbit interaction of conduction electrons in one-dimensional (1D) TI nanostructures.[16-18] Aharonov–Bohm (AB) conductance oscillations, characterized by the oscillation period of a single flux quantum $\Phi_0 = h/e$, have also been observed for TI nanoribbons and nanowires.[20-23] The AB oscillations in mesoscopic ring geometry are usually explained by the interference between partial waves encircling the magnetic flux enclosed by the electron paths. However, the AB oscillations from surface electrons in nanoribbons or nanowires under axial magnetic fields can be understood more properly in terms of the formation of 1D subbands, quantized along the circumference (with integer indices $n$) and modulated by the magnetic flux $\Phi$ threading the cross-sectional area of the quasi-1D nanostructure.[24-27] The dispersion relation for the 1D subbands in the TSS of TIs, as a function of the momentum $k$ along the nanowire, is as follows:[25, 27]

$$E(n,k,\Phi) = \pm h v_F \sqrt{\left(\frac{k}{2\pi}\right)^2 + \left(\frac{n+1/2-\Phi/\Phi_0}{L_p}\right)^2}, \qquad (1)$$

where $h$ is the Planck constant, $L_p$ is the perimeter length, and $v_F$ is the Fermi velocity that is assumed to be the same in both the azimuthal and axial directions. Electronic phase coherence is especially important for this 1D subband picture to hold, because the model explicitly relies on the existence of well-defined quantum eigenstates. Another important point is that the conductance maxima due to the AB oscillations can occur either for integer flux quanta ($\Phi = 0, \pm\Phi_0, \pm 2\Phi_0, \ldots$) or for half-integer flux quanta ($\Phi = \pm 1/2\Phi_0, \pm 3/2\Phi_0, \ldots$), depending on the location of the Fermi level and disorder strength in the nanowire.[26-27]

In this work, we have observed the phase-coherent quantum interference effects in magnetoconductance (MC) of devices made of $Bi_2Te_3$ single nanowires grown by using electrochemical deposition. The AB oscillations exhibiting both of the possible phases were clearly observed under magnetic fields parallel to the nanowire axis, verifying the existence of a highly coherent surface electron channel. The oscillation amplitude was found to decrease almost exponentially with increase in temperature $T$, implying that the phase coherence length $L_\varphi$ is inversely proportional to $T$. This suggests that the surface electron channel is in the quasi-ballistic regime. After carefully extracting the surface contribution from the MC measured in perpendicular magnetic fields, a WAL analysis was performed using a 1D localization theory. Our results demonstrate that the diameter-controlled $Bi_2Te_3$ nanowires, although synthesized by using a relatively coarse method of electrochemical deposition, exhibit a highly coherent surface electron channel that is essential for possible next-generation TI-based information technologies.

## 2. Experiments

Bi$_2$Te$_3$ nanowires were grown by electrochemical deposition by using porous anodic aluminum oxide membrane nanotemplates.[28] The cylindrical nanowires had a typical diameter and length of ~60 nm and 8−10 μm, respectively (Fig. 1(a)). Figure 1(b) is a scanning electron microscopy (SEM) image of the Bi$_2$Te$_3$ nanowires taken after removal of the AAO template. High-resolution transmission electron microscopy (HRTEM) and selected area electron diffraction (SAED) images show that the nanowires grow as single crystals in the [110] direction, normal to the *c*-axis (Fig. 1(c)). It is evidently shown that the nanowire is covered by a ~5-nm-thick native oxide layer, making the actual diameter of the nanowire conducting part ~50 nm. The X-ray diffraction (XRD) measurements on the nanowire array generate a strong peak from the (110) planes (Supplementary data Fig. S1), implying that the nanowires have a highly textured structure along the [110] direction, in accordance with the HRTEM result. For device fabrication, the nanowires were deposited onto highly *p*-doped silicon substrates, covered by a 300-nm-thick oxide layer. Metal (Pt or Ni) electrodes with a thickness of 100 nm are patterned by using standard electron beam lithography and electron beam evaporation to connect the nanowires to the Au bonding pads preformed on the substrates. The length *L* of the nanowire channel between the metal contacts was ~800 nm for all samples used in this study. A SEM image of a Bi$_2$Te$_3$ nanowire device (Pt1) is shown in Fig. 1(d). The electrical conductance *G* of single nanowire devices was measured by using a four-probe standard lock-in technique, using a physical property measurement system from Quantum Design, which provides a high-field (± 9 T) and low-temperature (2 K) environment for device measurements.

## 3. Results and discussion

Figure 2(a) shows the differential MC ($\Delta G \equiv G(B) - G(0)$) measured for the sample Pt2, with the magnetic field *B* applied parallel to the nanowire axis. It is apparent that conductance oscillations are superimposed on the slowly varying background MC drawn in solid curves, which is obtained by a fitting process. Each of the MC background curves consists of a polynomial and a Lorentzian, which has been used for convenience to simulate the conductance peak due to the WAL near *B* = 0 T. We note that the general shape of the MC background for $|B| > 1$ T does not change even when the magnetic field is applied perpendicular to the nanowire axis (Supplementary data Fig. S3). This suggests that the MC background is mainly determined by the magnetic field modulation of the electronic states in the nanowire bulk.[22]

The background MC has been subtracted from the MC data, and the residues denoted by $\delta G$ are shown in Fig. 2(b). It is clear that the residual MC is mainly comprised of a periodic oscillation. Figure 2(c) shows the fast Fourier transform (FFT) amplitudes of $\delta G$. A strong peak is observed at $1/B = 0.549$ T$^{-1}$, which corresponds to the oscillation period of $\Delta B = 1.82$ T. This oscillation period yields 54 nm as the diameter $d$, according to the relation for AB oscillations in a nanowire with a circular cross-section, $\pi(d/2)^2 \Delta B = \Phi_0$. This value of $d$ exactly matches the actual diameter of the conducting part of the nanowire deduced from the external diameter (64 nm) measured by using atomic force microscopy (AFM), and the oxide shell thickness (~5 nm) obtained from the HRTEM result (Fig. 1(c)). Besides the AB oscillations, the Altshuler−Aronov−Spivak (AAS) oscillations with a period of $h/2e$,[29] caused by the weak localization effect around the circumference, are also expected in a phase-coherent cylindrical surface electron system in the diffusive limit. However, Fig. 2(c) does not show a significant Fourier component of periodicity $h/2e$. It has been demonstrated that the AAS oscillations in TSS of TI can be easily disrupted by weak perpendicular magnetic fields.[27] It is possible that the absence of the AAS oscillations in our sample is caused by a slight misalignment of the applied field direction.

Figure 2(d) shows the AB oscillation amplitude ($\delta G_{AB}$) as a function of $T$. Evidently the temperature dependence does not follow the $\sim T^{-1/2}$ relation,[29] which is attributed to thermal averaging in diffusive systems over energy scales larger than the correlation energy $E_c$, that is, for $k_B T \gg E_c$. Depending on the type of system, the correlation energy could correspond to either the mean difference between the energy levels in the system or the Thouless energy, a scale of energy difference required for dephasing during the diffusion time through the system. It is known that the amplitude of AB oscillations in (quasi-)ballistic 1D rings is influenced by phase breaking and thermal averaging, each of which exhibits distinct temperature dependence.[30] Phase breaking along the circumference reduces the amplitude as $\sim \exp(-L_p/L_\varphi(T))$, where $L_\varphi$ ($\sim T^{-1}$) is the length of dephasing by inelastic scattering.[23, 30] Notice that in Fig. 2(d) with a semilog scale, the $\delta G_{AB}$ data points are almost linearly arranged, thus supporting the approximate relation $L_\varphi \propto T^{-1}$.

More accurate description of the thermal averaging effect can be obtained by the following equation,[27, 30] in which the zero-temperature conductance $G$ is convoluted with the derivative of the Fermi distribution function $f$.

$$G_{\text{avg}}(E_F, T, B) = \int G(E, 0, B) \left( -\frac{\partial f(E, E_F, T)}{\partial E} \right) dE . \qquad (2)$$

In its original context with a ring structure, $G$ in the above equation is meant to reflect the different phase gain dependent on $E$ along the electron paths. However, considering the 1D band model along the nanowire length, here $G$ can be put alternatively as $\sim\sin(2\pi E/\Delta + C(B))$, where $\Delta = hv_F/L_p$ is the inter-subband gap and $C(B)$ is a phase factor from the magnetic flux threading the nanowire. This substitution reflects the periodicity in the 1D density of states (DOS) with respect to $E$ and $B$, and results in the following relation for the AB oscillation amplitude,[27] $\delta G_{AB} \propto (2\pi^2 k_B T/\Delta)/\sinh(2\pi^2 k_B T/\Delta)$, reminiscent of that for the Shubnikov–de Haas oscillations. Here, $k_B$ is the Boltzmann constant. The above factors describing the influences of phase breaking and thermal averaging on $\delta G_{AB}$ are multiplicatively combined, and used for the fitting in Fig. 2(d). Using $L_p$ = 170 nm and $v_F \cong 3.7\times10^5$ m/s for the TSS of $Bi_2Te_3$,[11, 21] we obtain $L_\varphi$ = 950 nm at $T$ = 1 K. We note that $L_\varphi$ becomes shorter than $L_p$ at $T$ = 10 K, which explains the significant suppression of $\delta G_{AB}$ at the temperature.

Figure 2(e) shows the MC for another sample (Ni1) with Al/Ni contacts, measured at 2 K. $\delta G$ and its FFT amplitude have been obtained by using the same method as with sample Pt1, and clear AB oscillations are again observed after subtracting the background (Fig. 2(f)). The period of the AB oscillations for this sample is 2.3 T (Fig. 2(g)), giving an inner diameter of 48 nm. This value is also consistent with the nanowire height of 60 nm as measured by using AFM, given the oxide thickness of ~6 nm. A noticeable difference between Ni1 and Pt2 is that in the former sample, $\delta G$ exhibits a minimum at $B$ = 0 T, while in the latter, $\delta G$ exhibits a maximum at zero field. It has been predicted[26, 31] that in a TI nanowire, the phase of the AB oscillations, i.e., whether the oscillations exhibit a conductance maximum or minimum at zero flux, is determined by the doping level or by the relative location of the Fermi level in the TI. Recent work by Hong et al.[27] has provided experimental evidence for the above prediction by reporting gate-modulation of the AB oscillation phase in $Bi_2Se_3$ nanowires. Our data clearly show the MC traces that exhibit both phases of the AB oscillations in the $Bi_2Te_3$ nanowires, and support the interpretation of the AB oscillations as a DOS effect of the 1D subband.[27] In our case, the phase difference between Ni1 and Pt2 may be attributed to the differences between the nanowire diameters (leading to different inter-subband gaps) and the contact metals.

To further investigate the characteristics of the Bi$_2$Te$_3$ nanowire TSS, the MC of Pt2 has been measured in perpendicular magnetic fields. The perpendicular field MC exhibits a WAL conductance peak near $B = 0$ T, which is stronger and narrower than the peak with the parallel field in Fig. 2(a). Comparison between the parallel and perpendicular field MC suggests that only ~20% of the conductance enhancement near zero field under perpendicular fields can be attributed to the nanowire TSS, ascribing the rest to the nanowire bulk with strong spin−orbit coupling (Section 3, Supplementary data). Hence, extra care has been taken to extract the contribution of the TSS to WAL by using the two sets of the MC data. First, the AB oscillations were deducted from the parallel field MC. Then, the magnetic field value in the resultant MC data was scaled down by a factor of 3/4 (see Section 4 of Supplementary data for details). Finally, the rescaled MC, which presumably equals the WAL signal expected solely from the bulk electrons under perpendicular fields, was subtracted from the perpendicular field MC. The result of the above procedure, as shown in Fig. 3(a), can be regarded as the WAL signal from the TSS.

For a WAL analysis of the MC data, we have used the quasi-1D localization formula presented by Cha et al.,[17] neglecting the spin−spin scattering term by magnetic impurities. The 1D localization formula describing the quantum correction $\Delta G$ to the conductance is as follows:

$$\Delta G = \frac{\sqrt{2}e^2}{\pi\hbar}\frac{L_N}{L}\left[\frac{3}{2}\frac{\mathrm{Ai}\left(\frac{2L_N^2}{L_1^2}\right)}{\mathrm{Ai}'\left(\frac{2L_N^2}{L_1^2}\right)} - \frac{1}{2}\frac{\mathrm{Ai}\left(\frac{2L_N^2}{L_2^2}\right)}{\mathrm{Ai}'\left(\frac{2L_N^2}{L_2^2}\right)}\right], \qquad (3)$$

$$L_1 = \left(\frac{1}{L_\varphi^2} + \frac{4}{3L_{so}^2} + C\left(\frac{eWB}{\hbar}\right)^2\right)^{-1/2}, \quad L_2 = \left(\frac{1}{L_\varphi^2} + C\left(\frac{eWB}{\hbar}\right)^2\right)^{-1/2},$$

where $\hbar$ is the reduced Planck constant, and $W$ is the nanowire channel diameter. Ai is the Airy function, and Ai′ is its derivative. Here, $C$ is a numerical factor determined from algebraic calculations,[32] which is set to 2/9 instead of the usual value of 1/3, to adapt to the cylindrical geometry of the nanowire in our case (Section 4 of Supplementary data). The fitting parameters $L_\varphi$, $L_N$, and $L_{so}$ are the phase coherence length, the Nyquist length,[33-34] and the spin−orbit scattering length, respectively. Here, $L_\varphi$ reflects only the contributions from phase breaking mechanisms other than the Nyquist dephasing, which is caused by electron−electron interaction with small energy transfer (electric field fluctuations) and thus can be dominant in low-dimensional systems. From the curve fitting in Fig. 3, the following characteristic lengths were obtained for Pt2 at 2 K: $L_\varphi = 460$ nm, $L_N = 104$ nm, and $L_{so} = 105$ nm. Figure 3(b) shows that $L_N$ decreases with increasing $T$ as ~ $T^{-0.3}$, which conforms to the relation $L_N \sim T^{-1/3}$ theoretically expected for 1D systems[33, 35] and experimentally observed for similar TI nanowires.[17-18] $L_\varphi$ also decreases with increasing $T$ as ~ $T^{-1}$, in accordance with the analysis result of the AB oscillation amplitude given above.

Because the Nyquist length is a length scale by which the electron coherence is limited owing to the electric field fluctuations, it is subject to external factors such as electromagnetic noise[33] from the experimental setup. Therefore, we focus on the temperature dependence of $L_\varphi$. Unlike some earlier reports on the AB oscillations in TI nanoribbons that showed $\delta G_{AB}$ ~ $T^{-1/2}$,[20-21] our results indicate that $\delta G_{AB} \sim \exp(-L_p/L_\varphi(T))$ and $L_\varphi \propto T^{-1}$, with the inelastic dephasing time $\tau_\varphi$ inversely proportional to $L_\varphi$ as in ballistic systems. This temperature dependence implies that the TSS in our sample is in the quasi-ballistic regime[23] along the circumferential direction. Such temperature dependence of $L_\varphi$ has been reported in various mesoscopic systems[23, 30, 36-37] and attributed to the coupling to the environmental charge fluctuations.[38] This suggests that there can be room for enhancement of $L_\varphi$, possibly through control of nanowire doping or by modification of the device structure. Incidentally, with the quasi-ballisticity in mind, it is natural to question the applicability of the 1D localization theory. It is known that ballistic motion of electrons cause the flux cancellation effect, which reduces the effect of magnetic fields on weak localization/antilocalization.[39] Considering the functional form of Eq. (3), it is likely that $L_N$ is underestimated and $L_{so}$ is overestimated by the fitting. As a result, the actual spin−orbit coupling strength in the TSS can be stronger than what is deduced from Fig. 3(b).

## 4. Conclusion

Synthesis of semiconducting nanowires by electrochemical deposition has its advantages compared with other methods such as chemical vapor deposition. Although it may not be easy to achieve a high level of chemical purity and structural/stoichiometric perfection by using electrochemical deposition, the synthesis can be performed with exceedingly high throughput and relatively low installation costs by using the method. In particular, the method offers an almost perfect control of the nanowire size enabled by the choice of template. Our results show that the electrochemically synthesized and diameter-controlled $Bi_2Te_3$ nanowires are capable of highly coherent electrical conduction through the TSS, which constitutes the potential of TIs for future spintronics and quantum computing applications.

**Acknowledgement**

This work was supported by a Korea University Grant. We are grateful to Jung-Won Chang for useful discussions.

**Figure Captions**

**Figure 1.** (a) A top-view scanning electron microscopy (SEM) image of $Bi_2Te_3$ nanowires grown in an AAO template. (b) A side-view SEM image of the $Bi_2Te_3$ nanowires of a larger mean diameter, taken after removal of the AAO template. (c) An HRTEM image and a SAED pattern (inset) obtained from a $Bi_2Te_3$ nanowire. A ~5-nm-thick oxide shell is visible on the surface. (d) An SEM image of a $Bi_2Te_3$ nanowire device (Pt1) with four-probe Pt contacts. The channel length is 800 nm.

**Figure 2.** (a) The MC data (symbols) for the sample Pt2 under axial magnetic fields, at four different temperatures, ranging from 2 K to 10 K. Each MC background (solid curves) consists of a Lorentzian and a polynomial centered at $B = 0$ T. (b) Residual MC ($\delta G$) obtained by subtracting the background MC from the MC data. Periodic oscillations of $\delta G$ are clearly observed. The black curves are the fitting results by the cosine function. The curves are offset vertically for visual clarity. (c) FFT amplitude spectra of $\delta G$ with different temperatures. (d) Temperature dependence of the AB oscillation amplitude $\delta G_{AB}$. The solid curve is the result of the fitting discussed in the main text, and the dotted curve is drawn for comparison with $\sim T^{-1/2}$ dependence. (e) MC data from another sample (Ni1) with a smaller diameter. (f) Residual MC ($\delta G$) for Ni1. (g) FFT amplitude spectrum for Ni1. AB oscillations for Ni1 exhibit a phase opposite to that for Pt2.

**Figure 3.** (a) The perpendicular field MC data processed to include only the contribution from the surface channel of the sample Pt2. The solid curves are the results of the WAL fitting using the 1D localization formula of Eq. (3). The curves are offset vertically for visual clarity. (b) The phase coherence length $L_\varphi$, the Nyquist length $L_N$, and the spin−orbit scattering length $L_{so}$, as functions of temperature obtained from the fitting in (a). Temperature dependence of $L_\varphi$ and $L_N$ is fitted with $\sim T^{-\alpha}$. The exponent $\alpha$ is ~1 for $L_\varphi$ (black solid line) and ~0.3 for $L_N$ (red dashed line).

# References

[1] M.Z. Hasan, C.L. Kane, Colloquium: Topological insulators, Reviews of Modern Physics, 82 (2010) 3045-3067.

[2] X.-L. Qi, S.-C. Zhang, Topological insulators and superconductors, Reviews of Modern Physics, 83 (2011) 1057-1110.

[3] B.A. Bernevig, T.L. Hughes, S.C. Zhang, Quantum spin Hall effect and topological phase transition in HgTe quantum wells, Science, 314 (2006) 1757-1761.

[4] L. Fu, C. Kane, E. Mele, Topological Insulators in Three Dimensions, Physical Review Letters, 98 (2007) 106803.

[5] M. Konig, S. Wiedmann, C. Brune, A. Roth, H. Buhmann, L.W. Molenkamp, X.L. Qi, S.C. Zhang, Quantum spin hall insulator state in HgTe quantum wells, Science, 318 (2007) 766-770.

[6] D. Hsieh, D. Qian, L. Wray, Y. Xia, Y.S. Hor, R.J. Cava, M.Z. Hasan, A topological Dirac insulator in a quantum spin Hall phase, Nature, 452 (2008) 970-974.

[7] Y.L. Chen, J.G. Analytis, J.H. Chu, Z.K. Liu, S.K. Mo, X.L. Qi, H.J. Zhang, D.H. Lu, X. Dai, Z. Fang, S.C. Zhang, I.R. Fisher, Z. Hussain, Z.X. Shen, Experimental realization of a three-dimensional topological insulator, Bi2Te3, Science, 325 (2009) 178-181.

[8] D. Hsieh, Y. Xia, D. Qian, L. Wray, J.H. Dil, F. Meier, J. Osterwalder, L. Patthey, J.G. Checkelsky, N.P. Ong, A.V. Fedorov, H. Lin, A. Bansil, D. Grauer, Y.S. Hor, R.J. Cava, M.Z. Hasan, A tunable topological insulator in the spin helical Dirac transport regime, Nature, 460 (2009) 1101-1105.

[9] Y. Xia, D. Qian, D. Hsieh, L. Wray, A. Pal, H. Lin, A. Bansil, D. Grauer, Y.S. Hor, R.J. Cava, M.Z. Hasan, Observation of a large-gap topological-insulator class with a single Dirac cone on the surface, Nature Physics, 5 (2009) 398-402.

[10] D. Hsieh, Y. Xia, L. Wray, D. Qian, A. Pal, J.H. Dil, J. Osterwalder, F. Meier, G. Bihlmayer, C.L. Kane, Y.S. Hor, R.J. Cava, M.Z. Hasan, Observation of Unconventional Quantum Spin Textures in Topological Insulators, Science, 323 (2009) 919-922.

[11] D.X. Qu, Y.S. Hor, J. Xiong, R.J. Cava, N.P. Ong, Quantum oscillations and hall anomaly of surface states in the topological insulator Bi2Te3, Science, 329 (2010) 821-824.

[12] Z. Ren, A.A. Taskin, S. Sasaki, K. Segawa, Y. Ando, Large bulk resistivity and surface quantum oscillations in the topological insulator Bi_{2}Te_{2}Se, Physical Review B, 82 (2010) 241306.

[13] Z. Wang, T. Lin, P. Wei, X. Liu, R. Dumas, K. Liu, J. Shi, Tuning carrier type and density in Bi[sub 2]Se[sub 3] by Ca-doping, Applied Physics Letters, 97 (2010) 042112.

[14] H. Steinberg, D.R. Gardner, Y.S. Lee, P. Jarillo-Herrero, Surface State Transport and Ambipolar Electric Field Effect in Bi2Se3 Nanodevices, Nano Lett, 10 (2010) 5032-5036.

[15] J.G. Checkelsky, Y.S. Hor, R.J. Cava, N.P. Ong, Bulk Band Gap and Surface State

Conduction Observed in Voltage-Tuned Crystals of the Topological Insulator Bi_{2}Se_{3}, Physical Review Letters, 106 (2011) 196801.

[16] L.D. Alegria, M.D. Schroer, A. Chatterjee, G.R. Poirier, M. Pretko, S.K. Patel, J.R. Petta, Structural and electrical characterization of Bi(2)Se(3) nanostructures grown by metal-organic chemical vapor deposition, Nano Lett, 12 (2012) 4711-4714.

[17] J.J. Cha, M. Claassen, D. Kong, S.S. Hong, K.J. Koski, X.L. Qi, Y. Cui, Effects of magnetic doping on weak antilocalization in narrow Bi2Se3 nanoribbons, Nano Lett, 12 (2012) 4355-4359.

[18] W. Ning, H. Du, F. Kong, J. Yang, Y. Han, M. Tian, Y. Zhang, One-dimensional weak antilocalization in single-crystal Bi2Te3 nanowires, Scientific reports, 3 (2013) 1564.

[19] G. Bergmann, Weak localization in thin films a time-of-flight experiment with conduction electrons, Phys. Rep., 107 (1984) 1.

[20] H.L. Peng, K.J. Lai, D.S. Kong, S. Meister, Y.L. Chen, X.L. Qi, S.C. Zhang, Z.X. Shen, Y. Cui, Aharonov-Bohm interference in topological insulator nanoribbons, Nature Materials, 9 (2010) 225-229.

[21] F.X. Xiu, L.A. He, Y. Wang, L.N. Cheng, L.T. Chang, M.R. Lang, G.A. Huang, X.F. Kou, Y. Zhou, X.W. Jiang, Z.G. Chen, J. Zou, A. Shailos, K.L. Wang, Manipulating surface states in topological insulator nanoribbons, Nature Nanotechnology, 6 (2011) 216-221.

[22] M. Tian, W. Ning, Z. Qu, H. Du, J. Wang, Y. Zhang, Dual evidence of surface Dirac states in thin cylindrical topological insulator Bi(2)Te(3) nanowires, Scientific reports, 3 (2013) 1212.

[23] J. Dufouleur, L. Veyrat, A. Teichgräber, S. Neuhaus, C. Nowka, S. Hampel, J. Cayssol, J. Schumann, B. Eichler, O.G. Schmidt, B. Büchner, R. Giraud, Quasiballistic Transport of Dirac Fermions in a Bi2Se 3 Nanowire, Physical Review Letters, 110 (2013) 186806.

[24] J. Minkyung, S.L. Joon, S. Woon, H.K. Young, D.L. Sang, K. Nam, P. Jeunghee, M.S. Choi, S. Katsumoto, L. Hyoyoung, K. Jinhee, Quantum Interference in Radial Heterostructure Nanowires, Nano Lett, 8 (2008) 3189-3193.

[25] R. Egger, A. Zazunov, A.L. Yeyati, Helical Luttinger liquid in topological insulator nanowires, Physical Review Letters, 105 (2010) 136403.

[26] J.H. Bardarson, P.W. Brouwer, J.E. Moore, Aharonov-Bohm oscillations in disordered topological insulator nanowires, Physical Review Letters, 105 (2010) 156803.

[27] S.S. Hong, Y. Zhang, J.J. Cha, X.L. Qi, Y. Cui, One-dimensional helical transport in topological insulator nanowire interferometers, Nano Lett., 14 (2014) 2815-2821.

[28] H.S. Shin, S.G. Jeon, J. Yu, Y.S. Kim, H.M. Park, J.Y. Song, Twin-driven thermoelectric figure-of-merit enhancement of Bi 2Te3 nanowires, Nanoscale, 6 (2014) 6158-6165.

[29] S. Washburn, R.A. Webb, AHARONOV-BOHM EFFECT IN NORMAL METAL QUANTUM COHERENCE AND TRANSPORT, Advances in Physics, 35 (1986) 375-422.

[30] A.E. Hansen, A. Kristensen, S. Pedersen, C.B. Sørensen, P.E. Lindelof, Mesoscopic

decoherence in Aharonov-Bohm rings, Physical Review B - Condensed Matter and Materials Physics, 64 (2001) 045327.

[31] Y. Zhang, A. Vishwanath, Anomalous Aharonov-Bohm Conductance Oscillations from Topological Insulator Surface States, Physical Review Letters, 105 (2010) 206601.

[32] B.L. Altshuler, A.G. Aronov, MAGNETORESISTANCE OF THIN-FILMS AND OF WIRES IN A LONGITUDINAL MAGNETIC-FIELD, JETP Letters, 33 (1981) 499-501.

[33] B.L. Altshuler, A.G. Aronov, D.E. Khmelnitsky, Effects of electron-electron collisions with small energy transfers on quantum localisation, J. Phys. C : Solid State Phys., 15 (1982) 7367-7386.

[34] P. Echternach, M. Gershenson, H. Bozler, A. Bogdanov, B. Nilsson, Nyquist phase relaxation in one-dimensional metal films, Physical Review B, 48 (1993) 11516-11519.

[35] H.J. Zhang, C.X. Liu, X.L. Qi, X. Dai, Z. Fang, S.C. Zhang, Topological insulators in $Bi_2Se_3$, $Bi_2Te_3$ and $Sb_2Te_3$ with a single Dirac cone on the surface, Nature Physics, 5 (2009) 438-442.

[36] P. Roulleau, F. Portier, P. Roche, A. Cavanna, G. Faini, U. Gennser, D. Mailly, Direct measurement of the coherence length of edge states in the integer quantum Hall regime, Physical Review Letters, 100 (2008) 126802.

[37] K.-T. Lin, Y. Lin, C.C. Chi, J.C. Chen, T. Ueda, S. Komiyama, Temperature- and current-dependent dephasing in an Aharonov-Bohm ring, Physical Review B, 81 (2010) 035312.

[38] G. Seelig, M. Buttiker, Charge-fluctuation-induced dephasing in a gated mesoscopic interferometer, Physical Review B, 64 (2001) 245313.

[39] J.J. Lin, J.P. Bird, Recent experimental studies of electron dephasing in metal and semiconductor mesoscopic structures, Journal of Physics-Condensed Matter, 14 (2002) R501-R596.

**Figure 1.**

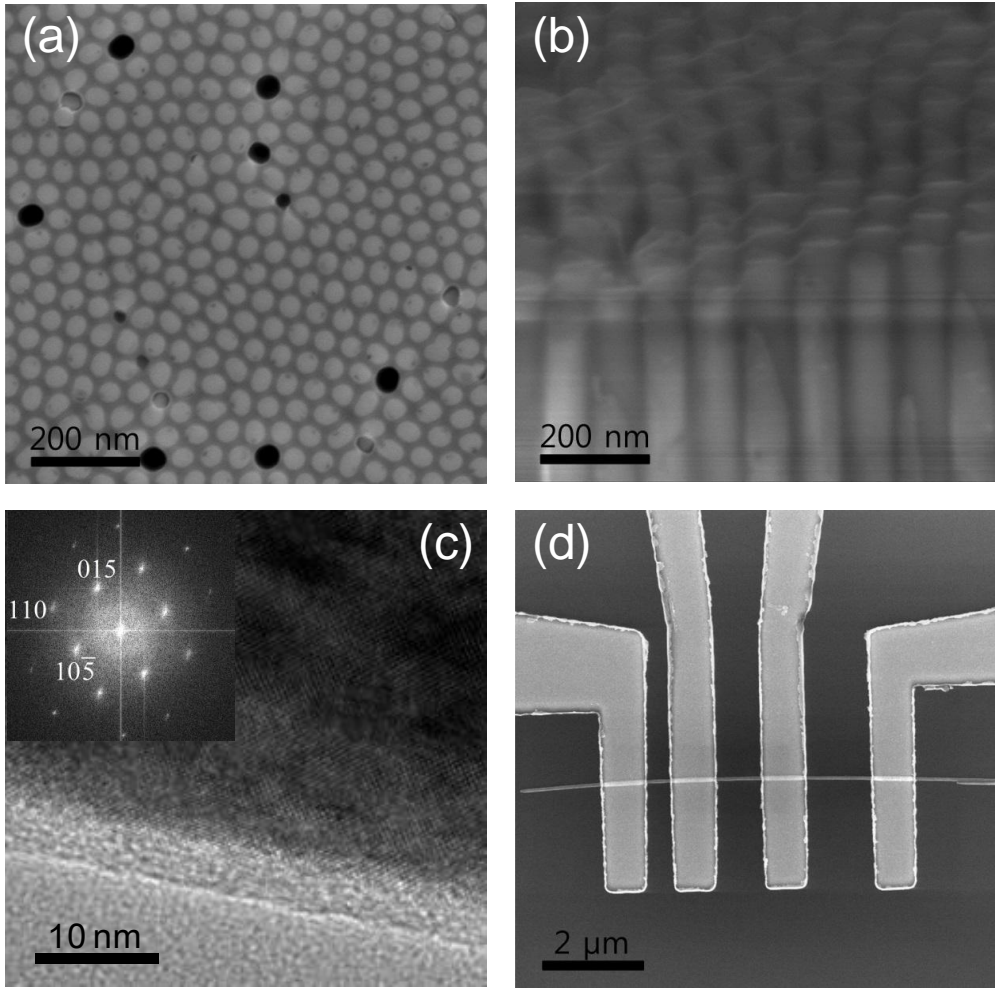

**Figure 2.**

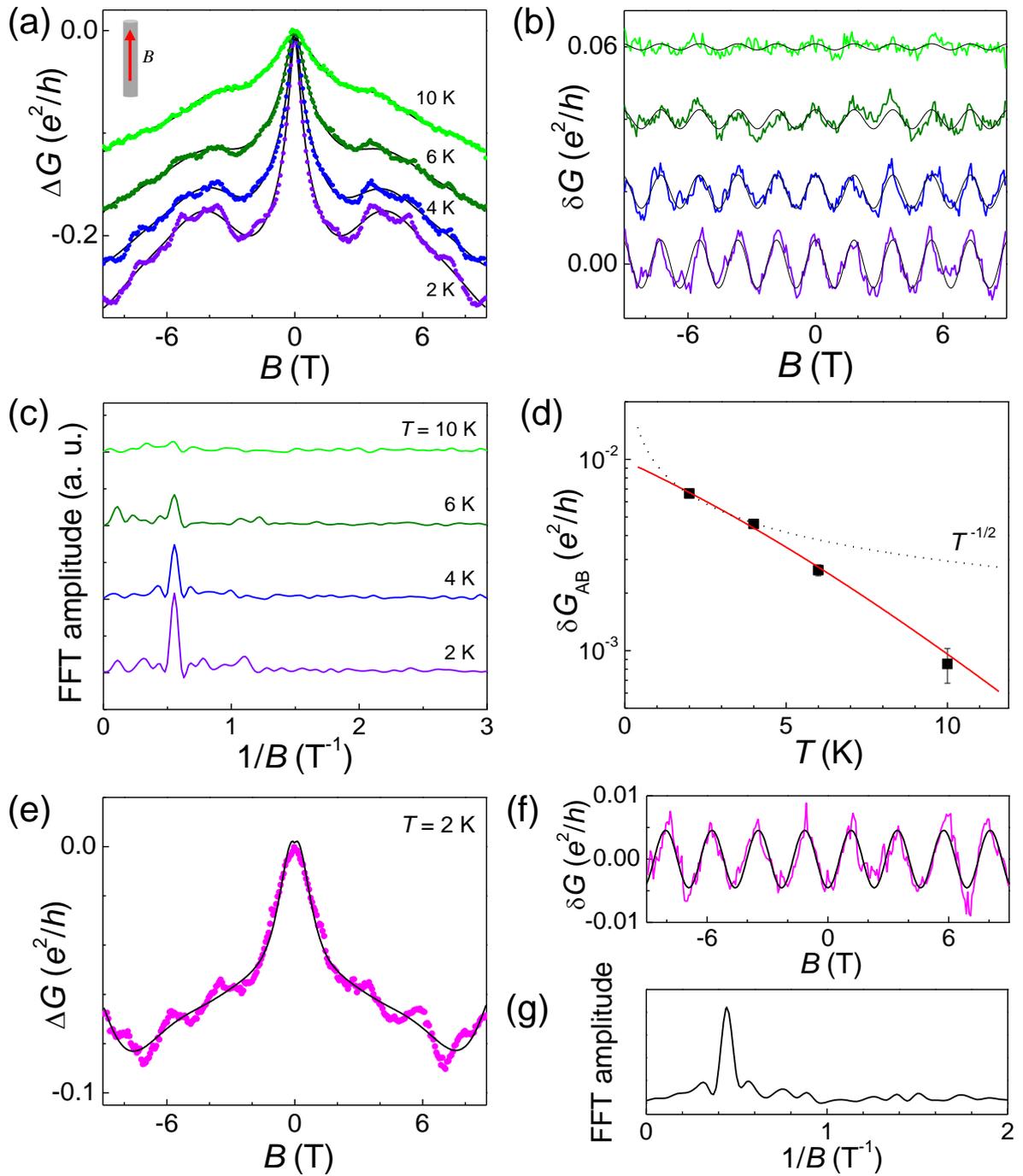

**Figure 3.**

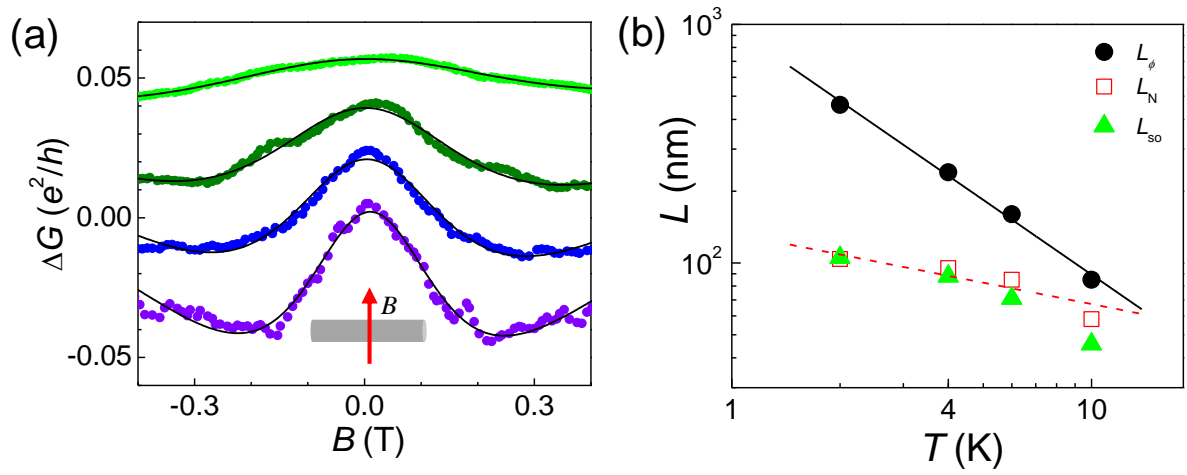